\documentstyle[prl,eqsecnum,aps,epsf,floats]{revtex} 
\begin{document} 

\draft 
\preprint{ } 
\title{Spin-charge separation in the single hole doped Mott antiferromagnet}
\author{Z.Y. Weng,$^a$ V.N. Muthukumar,$^b$ D.N. Sheng,$^{a,c}$ C.S. Ting$^a$} 
\address{$^a$Texas Center for Superconductivity, 
University of Houston, Houston, TX 77204-5506 }
\address{$^b$Joseph Henry Laboratories of Physics,
Princeton University, Princeton, NJ 08544}  
\address{$^c$Department of Physics and Astronomy, California State University,
Northridge, CA 91330}
\maketitle 
\date{today}
\begin{abstract} 
The motion of a single hole in a Mott antiferromagnet is investigated
based on the $t-J$ model. An exact expression of the energy spectrum is
obtained, in which the irreparable phase string effect [Phys. Rev. Lett.
{\bf 77}, 5102 (1996)] is explicitly present. By identifying the phase
string effect with spin backflow, we point out that spin-charge 
separation must exist in such a system: the doped hole has to decay into a 
neutral spinon and a spinless holon, together with the phase
string. We show that while the spinon remains coherent, 
the holon motion is deterred by the phase string, 
resulting in its localization in space.
We calculate
the electron spectral function which explains the line shape
of the spectral function as well as the ``quasiparticle'' spectrum observed in 
angle-resolved photoemission experiments. 
Other analytic and numerical approaches are discussed based 
on the present framework.

\end{abstract} 

\section{introduction}

Since the discovery of high-$T_c$ superconductivity, there has been 
much effort to elucidate
the properties of doped Mott insulators. In this context, the specific
case of one hole in a Mott insulating antiferromagnet (half filled band)
has been the subject of extensive theoretical and experimental
investigation. These studies address the basic question, whether
the motion of a hole in the antiferromagnet can possibly be described
within a quasiparticle approach. Photoemission spectroscopy provides 
valuable information that can help resolve this issue. Experimental 
results from angle resolved photoemission spectroscopy 
(ARPES) are now available for Sr$_2$CuO$_2$Cl$_2$ 
as well as for Ca$_2$CuO$_2$Cl$_2$ \cite{wells,t'-t'',ronning,add1,add2}. Both these materials
are Mott insulators and are parent compounds of the high-$T_c$ cuprates.
The results from ARPES can be summarized as follows: 
(i) the spectral features
observed are {\em not} sharp at all. Quite to the contrary, an 
intrinsic broad feature extending to energies of the order of 1.5 eV before
merging into the main valence band 
is seen. This is to be contrasted with sharp spectral features that one
expects in a quasiparticle scenario; (ii) the observed dispersion is
isotropic around ${\bf k}_0 = (\pi/2, \pi/2)$; (iii) 
the measurements of Ronning {\em et al.} \cite{ronning}
reveal the presence of a so-called remnant Fermi surface  of the 
momentum structure in the energy-integrated spectral function.

The broad spectral features seen in ARPES strongly suggest the breakdown
of the quasiparticle picture. Based on such considerations,
Laughlin \cite{laughlin} conjectured the failure of quasiparticle 
theory in these materials, and further proposed, that the observed 
isotropic dispersion has its origins in an underlying spinon spectrum.
This picture envisages spin-charge separation with the scale of the
observed dispersion determined by the superexchange $J$.

An entirely different picture is presented by the 
self-consistent Born approximation (SCBA) approach 
\cite{scba,kane,scba2,scba3}. Though this scheme is based on the
$t-J$ model, it depicts a spin-polaron picture for the single hole case 
where the doped hole behaves
not very different from a quasiparticle in the Landau-Fermi liquid theory. The 
SCBA results have also been supported by exact numerical calculations 
on finite lattices 
\cite{diagonalization} up to 32 sites \cite{leung} as well as
variational calculations on larger lattices\cite{lee}. Here, the consistency
amongst different numerical methods mostly concerns the spectrum 
$\epsilon_{\bf k}$, 
usually defined as the {\em minimum} energy for a given momentum ${\bf k}$.
The spectrum is {\em anisotropic} around the ``Fermi 
points'' ${\bf k}_0=(\pm \pi/2, \pm\pi/2)$ in all these calculations. It is
called a ``quasiparticle'' spectrum since a sharp peak in 
the spectral function usually appears at $\epsilon_{\bf k}$ in both 
the SCBA as well as results from
exact diagonalization. Note that these results are {\em not}
consistent with ARPES which exhibits an {\em isotropic} dispersion 
around ${\bf k_0}$, and rather
broad spectral features. Thus, to account for the former, the inclusion
of second ($t'$) and third ($t''$) nearest neighbor hoppings to the $t-J$ 
model have been proposed in the literature \cite{t'-t'',t',add,tohyama}.
However, this approach is meaningful only when the quasiparticle description 
holds, {\em i.e.}, only if a sharp
quasiparticle peak exists at $\epsilon_{\bf k}$ 
in the $t-J$ model. As illustrated in Fig. 1, if in the
thermodynamic limit, 
the spectral function for a given momentum ${\bf k}$ does not show a
sharp quasiparticle peak,
then, notwithstanding how accurately $\epsilon_{\bf k}$ is
determined by various theoretical methods, it has no experimental
\begin{figure}[ht!]
\epsfxsize=5.0 cm
\centerline{\epsffile{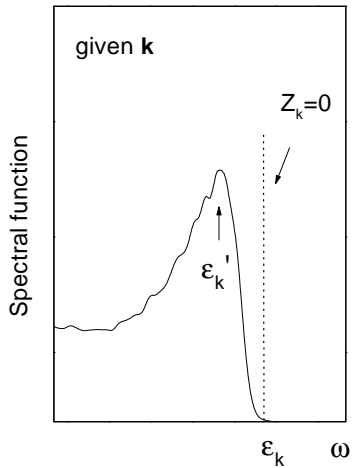}}
\vspace{2mm}
\caption{Schematic illustration of a spectral function at a given ${\bf k}$
in which the quasiparticle weight $Z_{\bf k}=0$ at the energy bottom $\epsilon_{\bf k}$. Here the ``quasiparticle'' peak at $\epsilon_{\bf k}'$ 
has nothing to do with the Landau-Fermi quasiparticle.} 
\label{fig:1}
\end{figure}
implication. For, in this case, the higher energy ${\epsilon}_{\bf k}'$ 
at the broad ``peak'' in Fig. 1, which is observable experimentally, 
may have nothing to do with the anisotropic $\epsilon_{\bf k}$.

On more general grounds, the breakdown of the Landau-Fermi quasiparticle 
picture has been discussed
by Anderson for a doped Mott insulator in the presence
of an upper Hubbard band \cite{book}. He argues that the 
quasiparticle weight $Z$
vanishes due to unrenormalizable Fermi-surface phase shifts 
induced when a particle is injected into the Mott insulator. Indeed, 
based on a rigorous formulation of the single-electron Green's function in 
the one-hole case using the $t-J$ model, 
it has been demonstrated\cite{string,string1} 
that the quasiparticle weight $Z$ at the Fermi surface must vanish in the 
thermodynamic limit due to the presence of the phase string effect. Such a
phase string effect can be considered as the equivalent of the 
phase shifts proposed by Anderson in the Hubbard model.

This paper concerns the effect of the aforementioned phase string on the
dynamics of a single hole introduced in an antiferromagnet. We show how
the phase string leads to the frustration of kinetic energy of the hole.
We then show that the phase string effect is related to spin-charge
separation and that the charge carrier is actually a spinless holon. We
find that the holon propagator is localized in space, owing to the phase
string. This is in contrast to the prediction of SCBA that the 
hole behaves as a Landau quasiparticle. 
Our result for the holon propagator
is consistent with an earlier conclusion 
that the quasiparticle weight is zero 
for the doped hole \cite{string,string1}. It does not 
necessarily contradict finite-size 
calculations, as the localization length scale turns out to be much larger
than typical sample sizes in numerical studies. 
We find the only coherent object to be a neutral spinon excitation 
created by the doped hole whose energy spectrum is 
responsible for an isotropic ``quasiparticle'' dispersion. 
Thus, our results provide a natural explanation of 
the ARPES results and support the conjecture made in Ref.\onlinecite{laughlin}.
We also discuss how the observed line shapes reflect spin-charge 
separation, and finally, in the Appendix, how the remnant Fermi surface 
structure may be understood as a peculiar consequence of the phase string effect 
at higher energies.

\section{Phase string: The key effect induced by the motion of a 
hole}

In this section, we examine the effects induced by the motion of a
single hole in an antiferromagnetic background. In Sec. IIA, we shall
review a few basic results of the slave-fermion formalism. 
We then go on to discuss the effect of the
phase string on the kinetic energy of the hole, spin-charge separation
and the localization of the holon due to the phase string.

\subsection{Slave-fermion formalism of the $t-J$ model}
We begin with 
the slave-fermion representation $c_{i\sigma}=f^{\dagger}_ib_{i\sigma}(-
\sigma)^i$\cite{remark1} and express the $t-J$ model 
$H_{t-J}=H_t + H_J$ in the following form:
\begin{equation}\label{hj}
H_J=-\frac J 2 \sum_{\langle ij\rangle} (\hat{\Delta}^s_{ij})^{\dagger}\hat{\Delta}^s_{ij}
\end{equation}
with 
\begin{equation}
\hat{\Delta}^s_{ij}=\sum_{\sigma}{b}_{i\sigma}{b}_{j-\sigma}~~,
\end{equation}
and
\begin{equation}\label{ht}
H_t=-t\sum_{\langle ij\rangle} 
\hat{B}_{ij} f^{\dagger}_if_j + h.c.~~,
\end{equation}
with 
\begin{equation}\label{b1}
\hat{B}_{ij}=\sum_{\sigma}
\sigma {b}^{\dagger}_{j\sigma}{b}_{i\sigma}~~.
\end{equation} 

At half-filling, the bosonic resonating-valence-bond (RVB) description given
by Liang, Doucot, and Anderson \cite{lda_88}
has provided by far, the most accurate picture 
(see Fig. 2) for both long-range as well as short-range 
antiferromagnetic (AF) correlations in two dimensions (2D). 
\begin{figure}[b!]
\epsfxsize=5.0 cm
\centerline{\epsffile{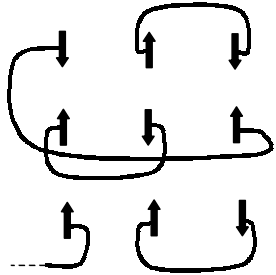}}
\vspace{2mm}
\caption{The resonating-valence-bond picture \protect \cite{lda_88,chen}
of bosonic spins provides
a highly accurate description of both short-range and long-range 
antiferromagnetic spin correlations in the Heisenberg model.
}
\label{fig:2}
\end{figure}
In this theory, it is found\cite{lda_88,chen} that the long-range AF 
correlations, including the AF long-range 
order (AFLRO) occurring in the thermodynamic limit, constitute only a {\em 
small} fraction of the ground-state energy and the dominant 
contribution mainly comes from the {\em short-range} RVB correlations. Thus, 
the ground state at half-filling may be regarded as AFLRO + RVB, 
of which, AFLRO
is the most vulnerable part easily removed by either temperature
or doping with little energy cost. The mean-field version of this bosonic RVB 
description is known as the Schwinger-boson mean-field theory 
(SBMFT)\cite{aa}. This theory  
works quite well at half-filling, and is characterized by a bosonic RVB
order parameter 
\begin{equation}\label{ds}
\Delta^s=\langle \hat{\Delta}^s_{ij}\rangle\neq 0~~.
\end{equation}

But away from half-filling, the problem is highly nontrivial. 
The reason can be attributed to the fact that the hopping integral 
$\langle \hat{B}_{ij}\rangle=0$ in $H_t$. This is generally true due to 
the sign $\sigma$ appearing in (\ref{b1}). So the motion of the hole
will be very sensitive to how the spin backflow $\hat{B}_{ij}$ is 
treated which is a non-perturbative problem and is the key issue to be
dealt with in the present work.  

In SCBA approach\cite{scba,kane,scba2,scba3}, the hole can
acquire some kinetic energy through the {\em dynamic fluctuations} of 
$\hat{B}_{ij}$. In this approach, only the long wavelength fluctuations 
associated with AFLRO is considered, where $\hat{B}_{ij}$ is approximated 
in large-$S$ (spin wave) expansion\cite{kane} by
\begin{equation}\label{b2}
\hat{B}_{ij}\approx \hat{B}_{ij}^{LSW}= b_0\sum_{\sigma}\sigma \left[{b}^{\dagger}_{j\sigma}+{b}_{i\sigma}\right],
\end{equation}
in which $b_0$ denotes the condensed part of the Schwinger boson field, 
corresponding to AFLRO \cite{remark2}. Now, the hopping of the
hole is assisted by the fluctuations of $\hat{B}^{LSW}$.  
This is the idea behind the SCBA, and 
within this approach 
it has been found that the hole behaves just like a 
Landau quasiparticle, known as the spin-polaron,
with four Fermi points at ${\bf k}_0 = (\pm \pi/2,\pm \pi/2)$
\cite{scba,kane,scba2,scba3}. 
We reemphasize that within 
this approximation only the {\em long-wavelength} fluctuations associated with
AFLRO are involved, and the dominant {\em short-range} RVB correlations, which 
are independent of AFLRO characterized by $b_0\neq 0$, are 
completely neglected.  

The validity of the SCBA approach is based on the presumption that
spin mismatches, described by the spin backflow
$\hat{B}_{ij}$ induced by the hopping of the hole, 
can be repaired through spin
flip process in $H_J$. Thus, the final state of the hole, after 
traversing a closed path through the antiferromagnetic background, is
identical to its initial state.
This would be true if one focuses only on, say, the 
$\hat{z}$-component of the spins. But since we are dealing with 
a quantum spin system in which each spin has three components 
that do not commute with one another, there will actually be three 
components in the spin mismatches induced
by the motion of the hole. 
For the spin-1/2 case, it has been explicitly proven\cite{string}
that such string-like spin defects {\it cannot} be simultaneously repaired
through $H_J$. Once the spin configuration ({\em i.e.}, the 
$\hat{z}$-component) disordered by the hole hopping is restored by spin 
flips at low energy, a string defect in the transverse components 
remains, which is described by a sequence of signs in the quantum description
of the hole. This defect is called the phase string\cite{string}.
The effect of this phase string, as we shall
see in the following two subsections, manifests itself both in the
expression for the kinetic energy of the hole as well as its propagator.

\subsection{Effect of the phase string on the energy of the hole} 

In this section, we shall demonstrate that the total energy at momentum 
${\bf k}$ for the one-hole case can be exactly formulated as 
\begin{equation}\label{ge}
E_{\bf k}=E_0 -\frac{t}{N}\sum_{ij}e^{i{\bf k}\cdot ({\bf r}_i-{\bf r}_j)}
M_{ij}~~,
\end{equation}
where
\begin{equation}\label{m}
M_{ij}\equiv \sum_{\{c\}\{\phi\}} M[c; \{\phi\}] 
(-1)^{N_c^{\downarrow}}~~.
\end{equation}
Here, $E_0$ denotes the spin ground-state energy with the hole being 
{\it fixed} at a given lattice site. The energy gain due to 
the hopping arises 
solely from the second term. 
In the expression for $M_{ij}$, the summations run
over all the possible paths 
of the hole connecting $i$ and $j$, $\{c\}$, as well as spin configurations, 
$\{\phi\}$. The weight functional $M$ is positive semi-definite, 
\begin{equation}\label{M}
M[c; \{\phi\}]\geq 0
\end{equation}
such that the phase factor $(-1)^{N_c^{\downarrow}}$ is 
``uncompensated'' in (\ref{m}). Here $N_c^{\downarrow}$ counts the 
total number of $\downarrow$ spins (or $\uparrow$ spins by symmetry) 
being {\it exchanged}
with the hole moving along the path $c$. As illustrated in Fig. 3, 
$(-1)^{N_c^{\downarrow}}= ... (+1)\times(-1)\times (-1)\times ...$ is 
the phase string on the path $c$, which was first 
identified in the exact formulation of the single-electron 
Green's function \cite{string}. 
Later we shall see that, owing to the factor
$(-1)^{N_c^{\downarrow}}$, the lowest energy of $E_{\bf k}$ is 
obtained for
${\bf k}={\bf k}_0$, consistent with numerical 
calculations\cite{diagonalization,leung,lee}.
\begin{figure}[b!]
\epsfxsize=7.0 cm
\centerline{\epsffile{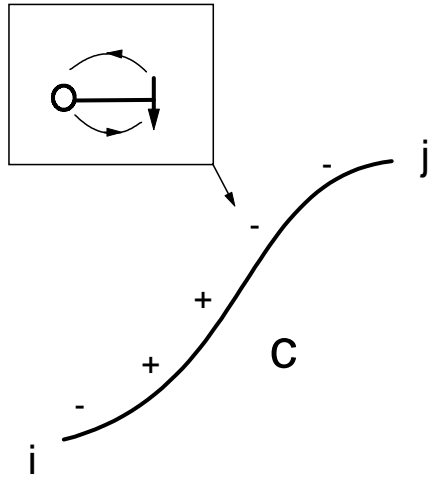}}
\vspace{2mm}
\caption{Phase string as a sequence of signs on the hole path $c$, determined 
by the index of each spin exchanged with the hole at every step of hopping. } 
\label{fig:3}
\end{figure}

From (\ref{ge}), (\ref{m}) and (\ref{M}), 
we come to an important conclusion. Without the phase string 
factor $(-1)^{N^{\downarrow}_c}$, 
the total energy would certainly be {\it lower} (as $M\geq 0$). Thus,
$(-1)^{N^{\downarrow}_c}$ represents 
the frustration on the kinetic energy of the hole. This effect is
(i) {\em irreparable} as no other signs can be
generated from the spin background to compensate it; 
(ii) {\em singular} since
a change of $N^{\downarrow}_c$ by $\pm 1$ will lead to a maximal 
change in $(-1)^{N^{\downarrow}_c}$; and (iii) thus expected 
to dominate the low energy dynamics of the hole. A similar effect
is well known in fermionic systems 
where each fermion's path is also weighted by a 
``phase string'' determined by
the number of fermions ``exchanged'' with it. 
The difference in this case is that the ``exchange'' 
is between two {\it different} species, the hole and spins.
Therefore, such a phase string effect implies an intrinsic 
{\it mutual statistics} between 
these two degrees of freedom \cite{string1}.

The proof of (\ref{ge})-(\ref{M}) is straightforward. We use the 
Wigner-Brillouin formula, 
\begin{equation}\label{expansion}
E_{\bf k}=E_0+\langle\Phi_0({\bf k})|
\left[H_t + H_t G_J (E_{\bf k}) H_t + ... \right]'
|\Phi_0({\bf k})\rangle~~,
\end{equation}
where $G_J(E)\equiv 1/(E-H_J)$ and $[...]'$ excludes 
$ |\Phi_0\rangle$ as the intermediate state. 
Here $ |\Phi_0({\bf k})\rangle=1/\sqrt{N}\sum_i
e^{i{\bf k}\cdot{\bf r}_i} |\Phi^{(i)}_0\rangle$ 
with $|\Phi^{(i)}_0\rangle$ denoting the ground state of 
$H_J$ with the hole 
localized at a site $i$, {\em viz.}, $H_J |\Phi^{(i)}_0\rangle =E_0 
|\Phi^{(i)}_0\rangle$. One can expand $|\Phi^{(i)}_0\rangle$ 
in terms of the complete spin-hole basis 
$\{|\phi; (i)\rangle\}$ ($\phi$ being a spin configuration with the hole
at site $i$) as 
$ |\Phi^{(i)}_0\rangle=\sum_{\phi} \chi^i_{\phi}|\phi;(i)\rangle$. 
As shown in Ref.\onlinecite{string,string1}, 
$\chi^i_{\phi}\geq 0$, which means that the Marshall
sign rule\cite{marshall} still applies to the doped ground state {\it if} the hole is not
moving. By inserting the complete set of basis states and following the
steps outlined in \cite{string} for calculating the single-hole
propagator, one can easily obtain (\ref{ge}), with the weight functional
\begin{equation}\label{wf}
M[{c};\{\phi\}] =\chi^i_{\phi}
\chi^j_{\phi'}\prod_{s=1}^{K-1}\langle\phi^{s+1};{(m_s)}|\hat{P}G_J(E_{\bf 
k})\hat{P}|\phi^{s};{(m_s)}\rangle(-t)~~,
\end{equation}
in which $\phi^s$ and $\phi^{s-1}$ correspond to spin configurations in 
intermediate states, 
and the hole site $m_s$ is on a path $c$ connecting two arbitrary sites 
$i$ and $j$: $m_0=i, m_1, ..., m_{K}=j $ ($\phi^0\equiv \phi$,
$\phi^{K}\equiv \phi'$). Following Ref.\onlinecite{string},
we can show that 
$\langle\phi^{s};{(m_s)}|\hat{P}G_J(E)\hat{P}|
\phi^{s-1};{(m_s)}\rangle \leq 0$ 
as long as $E<E_0$ (note that the projection operator $\hat{P}=1-
|\Psi_0\rangle\langle \Psi_0|$ has no effect in the thermodynamic limit 
$N\rightarrow \infty$). 
Since we are interested in low energy states
$E_{\bf k}<E_0$, the 
weight functional $M$ in (\ref{wf}) is always 
positive semi-definite. 

\subsection{Phase string effect and spin-charge separation}

In the above subsection, we showed that the phase string manifests
itself as irreparable sign (phase) frustrations induced by 
the hole motion. In this section, we shall establish an intrinsic
connection between such a phase string and spin-charge 
separation.

To this end, let us first consider the origin of the phase string factor 
$(-1)^{N_c^{\downarrow}}$ in (\ref{m}). Equation (\ref{b1}), 
defining $\hat{B}_{ij}$, can be rewritten as
\begin{equation}\label{b3}
\hat{B}_{ij}=\left(B^0_{ij}\right)e^{i\hat{\phi}_{ij}}~~,
\end{equation}
where 
\begin{equation}\label{b00}
B^0_{ij}=\sum_{\sigma}{b}^{\dagger}_{j\sigma}{b}_{i\sigma}~~,
\end{equation}
and
\begin{equation}\label{phi_ij}
\hat{\phi}_{ij}= \pm(\pi/2) [1 - \sigma_{ij}]~~.
\end{equation}
Here $\sigma_{ij}=1(-1)$, if an 
$\uparrow$ ($\downarrow$) spinon is 
{\it exchanged} with the holon at the link $(ij)$. For products around 
consecutive links enclosing a loop $\Gamma$,
\begin{equation}
\prod_{\Gamma}\hat{B}_{ij} \hat{B}_{jk} ...\hat{B}_{li}=
\left(\prod_{\Gamma}{B}^0_{ij}{B}^0_{jk} ...{B}^0_{li}\right)
e^{i\sum_{\Gamma}\hat{\phi}}~~,
\end{equation}
with
\begin{equation}
e^{i\sum_{\Gamma}\hat{\phi}}\equiv e^{i[\hat{\phi}_{ij}+\hat{\phi}_{jk}
+ ... +\hat{\phi}_{li}]}=(-1)^{N^{\downarrow}_{\Gamma}}~~,
\end{equation}
where $N^{\downarrow}_{\Gamma}$ denotes the total number of $\downarrow$ spins
{\em exchanged} with the hole along the loop $\Gamma$. Thus, the phase 
$\hat{\phi}_{ij}$ of the spinon backflow operator (\ref{b3}) is the source
of the phase string factor in (\ref{m}). This is illustrated by the inset in
Fig. 3.

Since the ground state satisfies the Marshall sign rule\cite{marshall}, it is 
easy to see that
\begin{equation}
\left\langle\prod_{\Gamma}{B}^0_{ij}{B}^0_{jk} 
...{B}^0_{li}\right\rangle_{\mbox{ half-filling}}> 0~~.
\end{equation}
(Note that the Marshall sign is totally gauged away 
in the definition of $b_{i\sigma}$ by the sign factor $(-\sigma)^i$ in the
slave-fermion decomposition\cite{string1}.)
This is of course consistent with the previous
conclusion that the nontrivial phases in the low-energy states all 
come from the phase string. A hole 
slowly hopping around the loop $\Gamma$ will then acquire a 
Berry's phase $\Phi_{\Gamma}$ given by
\begin{eqnarray}
\Phi_{\Gamma}&=&\mbox {
Im}\ln\left\langle\prod_{\Gamma}\hat{B}_{ij}\hat{B}_{jk} ...\hat{B}_{li}\right\rangle_{\mbox{ 
half-filling}}\nonumber\\
&=& \mbox {Im} \ln \left\langle e^{i\sum_{\Gamma}\hat{\phi}}\right\rangle_
{\mbox{half-filling}}.
\end{eqnarray}
So the phase string effect generally leads to a nontrivial Berry's 
phase picked up by the hole, as opposed to the quasiparticle picture 
(of SCBA, for instance) in which
the states of the hole before and after traversing a closed
path are identical.

Now that we have established the connection between
the {\em phase string effect} and the {\em spinon backflow}, 
we are in a position to discuss spin-charge separation. 
Let us initially suppose that the hole behaves like a quasiparticle 
with both charge and spin quantum numbers. This means that  
the holon and spinon should be confined together. In this case totally one
$\uparrow$ ($\downarrow$) spinon will have to be effectively
``transferred'' back from the final location of the hole to the initial 
location to ensure a precise spin quantum number $\downarrow$ ($\uparrow$)
being transported with the hole. It then requires that at each step of the holon
hopping, the spinon backflow is fully ${\em polarized}$, only to involve a 
$\uparrow$ ($\downarrow$) spinon being ``transferred'' backward. Otherwise,
since the spinon backflow or the phase string effect cannot be ``repaired'', any
local fluctuations of the spin polarization of the backflow spinons, no matter how weak they are, will be accumulated to become 
{\em arbitrarily} large at a sufficiently long path to {\em invalidate} that the 
hole carries a precise spin quantum number $s=1/2$. 
In other words, for the spin-charge confinement picture to hold, 
one must find that $N^{\downarrow}_c=0 $, or, $=$ the total number of links on
any path c, such that the phase string factor $(-1)^{N_c^{\downarrow}}$ becomes 
{\it trivial} no matter how long the path is. But based on (\ref{wf}) one can 
easily rule out such a scenario, which imposes the extreme restriction 
that the probability for any other choices of $N_c^{\downarrow}$
vanishes, since an equal probability for both kinds of spinon 
backflow is explicitly given in the hopping term (\ref{ht}). In fact, an
analysis\cite{string,string1} of the weight functional in the single-electron 
Green's function has already lead to the conclusion that the phase string effect 
must be {\em nontrivial} (which actually causes the spectral weight $Z$ vanish 
as shown in Refs.\onlinecite{string,string1}).

Therefore, {\em the irreparable nontrivial phase string effect directly leads to 
a true spin-charge separation in the single-hole doped Mott antiferromagnet}.
It confirms the conjecture made by Anderson\cite{book}. We note that
spin-charge separation has also been
discussed in some exact diagonalization 
approaches for momenta close to (0,$\pi$) \cite{add,tohyama}. 
In the following, we will go a step further and discuss some 
unique features associated with spin-charge separation.
  
Let us define the holon propagator $G_h$ in energy space, 
$G_h(E)\equiv\langle f_{j}G(E)f_{i}^{\dagger}\rangle_{\mbox{half-filling}}$, 
where $G(E)=1/(E-H_{t-J})$. On expanding in terms of $H_t$, 
$G(E)= G_J+ G_JH_tG_J+...$, we get
\begin{eqnarray}\label{gf0} 
G_h(j,i;E)&=&\sum_{\{c\}}\left\langle\left[\prod_{s=1}^{K}G_J(-t)\hat{B}_{m_sm_{s-1}}\right]G_J\right\rangle_{\mbox{half-filling}}\nonumber\\
&=& \sum_{\{c\}\{\sigma_s\}}
\left\langle\hat{S}_c(\{\sigma_s\})\right\rangle_{\mbox{half-filling}} 
(-1)^{N_c^{\downarrow}}~~,
\end{eqnarray}
where $\hat{S}_c (\{\sigma_s\})\equiv 
\left(\prod_{s=1}^{K}G_J(-t)\left[b^{\dagger}_{m_{s-
1}\sigma_{s}}b_{m_s\sigma_{s}}\right]\right)G_J$
with $N^{\downarrow}_c=\sum_s (1-\sigma_{s})/2$. In (\ref{gf0}),
$c$ 
denotes a path connecting two sites $i$ and $j$, and $m_s$ in 
the definition of $\hat{S}_c$ 
is a lattice site on any path $c$. As with the 
the positive semi-definite functional 
$M$ in the expression (\ref{m}) for $E_{\bf k}$, one can easily prove that 
$\langle\hat{S}_c\rangle\geq  0$ for energies $E<E_0$. 
This means that there are no other sources of phases 
at low energies to repair the phase string factor 
$(-1)^{N_c^{\downarrow}}$. It should be noted that
{\em the single-electron} Green's function for the one-hole case 
has been exactly formulated previously in a 
very similar fashion \cite{string}. 
There, it was found that each hole path is also modulated by the 
same phase string $(-1)^{N_c^{\downarrow}}$. 
This result taken in conjunction with (\ref{gf0}), 
shows that {\em the effect of the phase string is entirely in the holon sector}, 
while the spinon released from the hole is not directly associated with 
such phase frustrations. Such a picture is shown in Fig. 4. 
This is not entirely surprising, as we do not expect the motion of a
single hole to affect the (thermodynamically large) spin subsystem,
except possibly at short time scales or high energies.
\begin{figure}[t!]
\epsfxsize=7.0 cm
\centerline{\epsffile{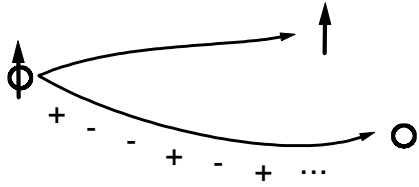}}
\vspace{2mm}
\caption{Spin-charge separation in the single-hole case: the injected hole 
decays into a neutral spinon and a spinless holon, and the holon will pick up 
the full effect of phase string. Representing the spinon backflow effect, the
irreparable phase string ensures spin-charge separation as discussed in the text. } 
\label{fig:4}
\end{figure}

\subsection {Localization of the holon due to the phase string effect }

Thus far, we have discussed the general properties of the phase 
string effect in a rigorous form. To further study the one-hole problem in
a quantitative way, we need a suitable framework for approximation. As 
mentioned earlier, the mean field version of the bosonic RVB theory (SBMFT) 
provides\cite{aa} a fairly good description of the undoped
antiferromagnet. We now assume, consistent with the discussion so far,
that the presence of one hole does not alter the low energy properties
of the spin subsystem as the irreparable phase string is to be solely
picked up by the hole. In the following we shall discuss a direct consequence of
the phase string effect on the charge part within this approximation: the
localization of the holon. 

At half-filling, SBMFT characterized by (\ref{ds}) is described 
by the Bogoliubov transformation\cite{aa}
\begin{equation}\label{bogo}
{b}_{i\sigma}=\frac 1 {\sqrt{N}}\sum_{\bf k}\eta_{{\bf k}\sigma} 
e^{i\sigma {\bf k}\cdot
{\bf r}_i} \left[u_{\bf k}\gamma_{{\bf k}\sigma}-v_{\bf k}\gamma_{{\bf k}-
\sigma}^{\dagger}\right]~~,
\end{equation}
where $\gamma_{{\bf k}\sigma}^{\dagger}$ is an operator that creates
an elementary spinon excitation with a gapless spectrum at $T=0^+$, 
\begin{equation}\label{sp}
E^s_{\bf k}=2.32 J\sqrt{1-\xi_{\bf k}^2},
\end{equation}
and $u_{\bf k}=1/\sqrt{2}({2.32J}/E^s_{\bf k}+1)^{1/2}$, $v_{\bf k}=1/\sqrt{2}({2.32J}/E^s_{\bf k}-1)^{1/2}sgn(\xi_{\bf k})$, with 
$\xi_{\bf k}=(\cos k_x +\cos k_y)/2$. Within this mean field theory, 
it is easy to verify that $\langle b^{\dagger}_{i\uparrow}b_{j\uparrow} 
\rangle = \langle b^{\dagger}_{i\downarrow}b_{j\downarrow} \rangle 
\equiv 0$.
If we use this result naively, when there are $N-1$ spins
(and a hole), we would conclude (erroneously) that $B^0=\langle B^0_{ij}
\rangle=0$, namely, the amplitude of the hopping integral vanishes. 
This is incorrect because the hopping term connects two spin subspaces, corresponding to the hole at site $i$ and $j$, respectively, which are
not necessarily identical in symmetry. This subtlety has to
be incorporated in any calculation, as will be done below.

As a first step, let us go back to the case of half filling.
Here, it is important to recognize that the factor $\eta_{\bf k\sigma}$ in
(\ref{bogo}) which satisfies $\eta_{\sigma}=\eta_{-\sigma}^*$ and $|\eta|=1$  
cannot be completely determined at the 
level of mean field theory; {\em i.e.}, the 
mean field order parameter $\langle {b}_{i\sigma}{b}_{j-\sigma}\rangle$ 
is independent of the choice of $\eta_{{\bf k}\sigma}$. Therefore, 
there is a hidden symmetry in SBMFT, which is related to the exact local 
particle-hole invariance of the system as to be discussed later.

We now exploit this symmetry in SBMFT for the case when there is one hole.
Fixing the position of the hole, one may define a 
subspace for the spins, which, at the level of mean field theory,
is again described by the 
Bogoliubov transformation (\ref{bogo}). 
It should be remembered that equation (\ref{bogo}) now describes an
$N-1$ spin subspace. By choosing $\eta_{\bf k\sigma}=
[-{\mbox sgn}(\xi_{\bf k})]^{k_h}$, 
where $k_h=0$ if the hole is 
on the even sublattice and $k_h =1$ if the hole is on the odd sublattice 
site, one has $\eta_{{\bf k}\sigma} \rightarrow -{\mbox 
sgn}(\xi_{\bf k}) \eta_{{\bf k}\sigma}$ when the hole hops
between the sublattices\cite{remark3}. Then noting that in $B^0_{ij}$, $b^{\dagger}_{j\sigma}$
and $b_{i\sigma}$ belong to two different spin subspaces, 
one obtains
\begin{equation}\label{b0}
{B}^0=\frac 2 N \sum_{{\bf k}\neq 0} |\xi_{\bf k}| v_{\bf k}^2
\approx 0.4~~.
\end{equation}
It is easy to verify that the above choice of $\eta_{{\bf k}\sigma} $
optimizes $B^0$.
Thus, we obtain a finite amplitude of the hopping integral for the motion 
of the holon between two sublattices. (Note that in the summation of (\ref{b0}), 
${\bf k}=0$ has been removed excluding the $b_0$ component. 
Using (\ref{b1}), one can easily see that $b_0$ part has no contribution 
due to the sign in $\sigma$.) The mean field result remains approximately the 
same for the superexchange term $H_J$ in which the holon position is fixed
and $\eta_{\bf k}$ plays no role.
 
This hopping amplitude $B_0\neq 0$ is consistent with
$\triangle^s \neq 0$, and we may understand this in the following way.
Physically, each spin subspace has a hidden local particle-hole 
symmetry, $b_{i\sigma} \rightarrow b^{\dagger}_{i-\sigma}$. It is exact and can 
be preserved in the RVB description if the no double occupancy constraint is 
strictly implemented 
locally. Let us assume that the holon is, say, at an odd sublattice site. Then,
on performing a transformation $b_{i\sigma} \rightarrow b^{\dagger}_{i-\sigma}$ only in the {\em corresponding spin 
subspace}, it is easy to see that $B^0_{ij}$ at the hopping bond 
transforms into the RVB order parameter $\triangle^s$. Note that at the mean-field level in SBMFT, the local hard-core constraint is relaxed where the particle-hole 
symmetry is no longer exact. But it can be checked that the symmetry $\eta_{{\bf k}\sigma} \rightarrow -{\mbox sgn}(\xi_{\bf k}) \eta_{{\bf k}\sigma}$ in SBMFT 
still approximately corresponds to
$b_{i\sigma} \rightarrow b^{\dagger}_{i-\sigma}$ at the global level according
to (\ref{bogo}) (It would be exact if $|u_{\bf k}|=|v_{\bf k}|$). It tells
us that the direct hopping term for the holon indeed originates from local RVB 
spin pairing with a particle-hole symmetry in each spin subspace.

With $B^0\neq 0$,
$G_h$ in the first line of (\ref{gf0}) may be written, 
in the limit of $E\rightarrow -\infty$, as
\begin{equation}
G_h({E\rightarrow -\infty})\approx 
\sum_{\{c\}}g_h^c\left\langle e^{i\sum_c\hat{\phi}}
\right\rangle_{\mbox{half-filling}}
\end{equation}
where $g_h^c= 2^{K}\langle \hat{S}_c\rangle_{\mbox{half-filling}}$. 
Note that in $E\rightarrow -\infty$ limit $G_J\rightarrow 1/E$ becomes 
commutable with $e^{i\hat{\phi}}$. We can now calculate the
phase string average $\langle e^{i\sum_c\hat{\phi}}\rangle_{\mbox{
half-filling}}$ explicitly as 
\begin{eqnarray}\label{string}
\left\langle e^{i\sum_c\hat{\phi}}\right\rangle_{\mbox{half-filling}} 
&=& e^{i\langle \sum_c\hat{\phi}\rangle}\langle e^{i
\sum_c\left(\hat{\phi}-\langle \hat{\phi}\rangle \right)}\rangle\nonumber\\
&=& e^{i{\bf k}_0\cdot ({\bf r}_i-{\bf r}_j)}e^{-
\frac{1}{2}\langle\left[\sum_c\left(\hat{\phi}-\langle 
\hat{\phi}\rangle\right)\right]^2
\rangle}~~,
\end{eqnarray}
where $\langle[\sum_c(\hat{\phi}-\langle \hat{\phi}\rangle)]^2
\rangle=\frac{\pi^2}{2}\langle\left(\sum^c_i :S^z_i:\right)^2\rangle$, 
and $:S^z_i:=S^z_i-\langle S^z_i\rangle$. Using the results for 
$\langle S^z_iS^z_j \rangle$ from SBMFT \cite{aa},
one finds in the large 
$|{\bf r}_{ij}|$ limit, the asymptotic form for
expression (\ref{string}) and thus, the
holon propagator as 
\begin{equation}\label{g_f} 
G_h(j,i;E)\sim  e^{i{\bf k}_0\cdot ({\bf r}_i-{\bf r}_j)} e^{-\frac{|{\bf 
r}_i-{\bf r}_j|} {\lambda_{L}}}~~.
\end{equation}
The localization length is determined numerically, 
for the case $E = -\infty$, as 
$\lambda_L(E=-\infty)\sim 2.2 a$ for ${\bf r}_{ij}$ parallel to the 
x- or y-axis (Here, $a$ is the lattice constant).

The exponential decay (\ref{g_f}) of the holon
propagator, {\em a typical characterization of localization phenomenon}, 
is fundamentally different from the power-law decay for the 
propagator generally expected when
the hole behaves like a well-defined 
quasiparticle. 
The very fact that the holon propagator behaves like (\ref{g_f}) 
as $E\rightarrow -\infty$ is enough to invalidate conventional 
quasiparticle behavior. 
As in conventional localization problems, we expect $\lambda_L(E)$ 
to increase 
with the energy $E$ and to represent the true localization length 
scale at 
$E\geq E_G$ ($E_G$ is the ground-state energy). 
Later, we shall determine the localization length 
$\lambda_L(E)$ numerically, in the physical regime of 
$E$. We find it to be generally larger than 
the overall scales of sample sizes ($<6a\times 6a $) 
used in exact-diagonalization calculations. 
So, the hole localization caused by the
phase string effect cannot be directly detected by 
exact diagonalization,
owing to the limitation of the sample size. 
On the contrary, as in conventional localization problems,
a localized electron can be mistaken for a delocalized electron
simply because the sample sizes are smaller than the localization
length. Thus, we argue that
the well defined quasiparticle peak seen in finite size 
calculations is an artifact of small sample sizes and cannot persist
at scales beyond $\lambda_L$.  

\section{Spectral function}
\subsection{Effective theory}
As pointed out earlier, the bosonic spin RVB 
pairing gives an extremely good description of the undoped antiferromagnet 
for {\em both} short-range and long-range spin-spin correlations. 
The direct hopping of the holon originates from {\em short-range} RVB pairing 
with a local particle-hole 
symmetry. Unlike the long-range spin correlations, the short-range 
correlations are not sensitive to doping and the local RVB order parameter
$\Delta^s$ provides a certain local ``rigidity'' 
that underpins the doped antiferromagnet, as long as the
the average spacing between the holes is
larger than the distance between nearest neighbors.
Indeed, such a direct hopping term can persist into the metallic 
(superconducting) phase at finite doping, as discussed in 
Ref. \onlinecite {string3}. 
However, unlike the metallic system, the one-hole case is different as
we do not expect the AF spin background to change at thermodynamic
scales. As discussed in the previous section, one may, to leading order,
assume the spin background to be the same as that at half-filling.
The feedback effect on the spin part at high-energy, 
short-distance scales will be considered in the Appendix.

In the previous section, we saw that the holon picks up the effect of
the phase string and that the effective Hamiltonian can be written as
\begin{equation}\label{hh}
H_h=-t_h\sum_{\langle ij\rangle} e^{i\hat{\phi}_{ij}}
f^{\dagger}_if_j+ h.c.
\end{equation}
with $t_h={B}^0t\simeq 0.4t$. Here, the nontrivial effect arises solely from
the phase $\hat{\phi}_{ij}$, reflecting the irreparable phase 
string effect. In terms of (\ref{phi_ij}), we may rewrite
\begin{eqnarray}
e^{i\hat{\phi}_{ij}}&=&e^{i{\bf k}_0\cdot ({\bf r}_i-{\bf r}_j)\left[1-\sigma_{ij}\right]}\nonumber\\
&\equiv & e^{i{\bf k}_0\cdot ({\bf r}_i-{\bf r}_j)}e^{-ia_{ij}}
\end{eqnarray}
where 
\begin{equation}
{\bf k}_0=\left(\pm\frac{\pi}{2a},\pm\frac{\pi}{2a}\right)
\end{equation}
and $a_{ij}={\bf k}_0\cdot({\bf r}_i-{\bf r}_j)\left[\sigma_{ij}\right]$.

To characterize the strength of $a_{ij}$, let us consider the gauge
invariant quantity $\sum_{\Box}a_{ij}$, {\em i.e.}, the fictitious ``magnetic'' 
flux seen by the holon hopping around a plaquette: $i\rightarrow 
i+\hat{x}\rightarrow
i+\hat{x}+\hat{y}\rightarrow i+\hat{y}\rightarrow i$. One can easily get
\begin{equation}\label{aflux}
\mbox{$\sum_{\Box}$}a_{ij}=\pm \pi\ \left(S^z_{i+\hat{x}}+ S^z_{i+\hat{x}+
\hat{y}}- S^z_{i+\hat{y}}-S^z_{i}\right).
\end{equation}
Generally, $\langle \sum_{\Box}a_{ij}\rangle=0$, and the strength of the 
quadratic fluctuations is given by
\begin{equation}\label{gamma1}
\left\langle \left({\mbox {$\sum_{\Box}$}}a_{ij}\right)^2\right\rangle= \pi^2\left(1-\frac 
4 3 \left\langle{\bf S}_i\cdot{\bf S}_{i+\hat{x}+\hat{y}}\right\rangle\right).
\end{equation}
Using the SBMFT one can estimate $\sqrt{\langle 
(\sum_{\Box}a_{ij})^2\rangle}\approx 0.86 \pi$ and find that, for two 
plaquettes separated far away from each other (i.e., the distance $R_{12}\gg 
a$),
\begin{equation}\label{gamma2}
\left\langle \left({\mbox {$\sum_{\Box}$}}a_{ij}\right)_1 \left({\mbox{$\sum_{\Box}
$}}a_{lk}\right)_2\right\rangle\sim O\left(\frac {a^4} {R_{12}^4}\right).
\end{equation}
Note that the spatial correlations between the fluxes threading through 
different plaquettes fall off rapidly in (\ref{gamma2}). This is because
the contribution from the long-range AF fluctuations is strongly canceled out
in (\ref{aflux}). Similarly, the correlations of fluxes per plaquettes at 
different times also decay very quickly. Thus, as the first order of 
approximation one may treat $a_{ij}$ as a gauge field describing random fluxes 
in the white-noise limit with the strength controlled by (\ref{gamma1}). 
(The phase string effect beyond the random flux approximation is briefly
discussed in the Appendix in conjunction with the momentum dependence of
the energy-integrated spectral function).

In the effective model (\ref{hh}), the effect from the longer-range spin 
correlations involving the AFLRO, which influence the hopping of the holon 
through (\ref{b2}), has been omitted. This process has been considered in
the SCBA approach as the {\em sole} source assisting the holon hopping. In the 
following we reexamine it in the presence of the bare 
holon term (\ref{hh}). With the Bose condensate $b_0\neq 0$ (AFLRO), 
one may express  
$c_{i\sigma}\equiv (-\sigma)^i\bar{c}_{i\sigma}+ 
(-\sigma)^if_i^{\dagger}:b_{i\sigma}:$ with 
$\bar{c}_{i\sigma}=b_0f^{\dagger}_i$ and 
$:b_{i\sigma}:= b_{i\sigma}-b_0$. Then, Eq.(\ref{b2}) leads to 
the following hopping Hamiltonian, 
\begin{equation}\label{lsw}
H_h^{LSW}=-t \sum_{\langle ij\rangle \sigma}
\sigma\left[\bar{c}_{i\sigma}:b^{\dagger}_{j
\sigma}:f_j+ f_{i}^{\dagger}:b_{i\sigma}:\bar{c}_{j\sigma}^{\dagger}\right]+ 
h.c.~~. 
\end{equation}
If there were no bare hopping (\ref{hh}) for the {\it holon}, $H_h^{LSW}$ 
would represent a virtual process, which, by the SCBA 
treatment\cite{scba,kane,scba2,scba3}, would give rise to 
the well-known quasiparticle description for 
$\bar{c}_{i\sigma}$. But in the presence of (\ref{hh}), $H_h^{LSW}$ 
literally describes the process for $\bar{c}_{\sigma}$ (the 
``quasiparticle'') to {\it decay} into a holon-spinon pair. In this case,
the propagator $\bar{G}_e$ for $\bar{c}_{\sigma}$ will no longer behave like
the one for a dressed quasiparticle found in SCBA. In contrast, to leading
order approximation, it will be simply proportional to the propagator of 
$f_i^{\dagger}:b_{i\sigma}:$ to be discussed below. So in the following we 
do not consider $H_h^{LSW}$ in (\ref{lsw}) and 
simply switch ${\bar c}_{i\sigma}$ off, by assuming either the sample size 
to be large but finite or $T=0^+$, such that $b_0=0$ without loss of generality.

\subsection{Spectral function}

In this subsection, we shall show that the main features observed in
ARPES can be described within our formalism. We attribute the broad
spectra that are observed to the fact that the physical electron is
a convolution of spinon and holon excitations. We show that the
dispersion of the photoelectron is governed by the dispersion of the
spinon, and consequently, isotropic. We also show that the phase string
plays a crucial role in causing the sharpest spectra at low binding energy
locating at $(\pm\pi/2, \pm\pi/2)$ ({\em i.e.}, the ``Fermi points'').

The single-electron propagator may be expressed in the 
following {\em decomposition} form 
\begin{equation}  \label{decomp}
G_e \approx i G_b \cdot G_h
\end{equation}
where 
\begin{equation}  \label{gb}
G_b (i,j;t)=-i(-\sigma)^{i-j}\left\langle T_t {b}_{i\sigma}(t){b}^{
\dagger}_{j\sigma}(0)\right\rangle,
\end{equation}
and
\begin{equation}  \label{gh}
G_h(i, j; t) =-i\left\langle T_t f^{\dagger}_i(t)f_j(0)\right\rangle.
\end{equation}
Here we mainly focus on the properties of the 
spectral function defined by $A^e({\bf k}, \omega)=-\frac 1 \pi
{\mbox {Im}}G_e({\bf k}, \omega){\mbox{sgn}}(\omega)$, which can be
measured by ARPES. Using the decomposition (\ref{decomp}), one finds 
the convolution law 
\begin{equation}\label{spectral}
A^e({\bf k},\omega)= \theta(-\omega)\frac 1 N \sum_{{\bf k}'}\int^0_{\omega}
d\omega' \rho_h({\bf k}'-{\bf k}, \omega'-\omega) \rho_b({\bf k}',\omega')
\end{equation}
at $T=0^+$ where $\rho_b$ and $\rho_h$ are the spectral function 
corresponding to $G_b$ and $G_h$, respectively.

By using (\ref{bogo}) one can easily obtain
\begin{equation}\label{spectralb}
{\mbox {Im}} G_b=-
\pi [u_{\bf k}^2\delta(\omega-E^s_{\bf k})-v_{\bf k}^2\delta(\omega+
E^s_{\bf k})].
\end{equation}
As noted above, in the one-hole case this
half-filling mean-field propagator should not be affected thermodynamically 
by the motion of the single holon.
Then the corresponding electron spectral function is written as
\begin{equation}\label{ae}
A^e({\bf k},\omega)=\frac 1 N \sum_{{\bf k}'} v_{{\bf k}'}^2 {\rho}_h({\bf
k}'-{\bf k}, -E^s_{{\bf k}'}-\omega)
\end{equation}   
where the condition $\rho_h\neq 0$ only at $\omega\geq 0$ is used which 
defines the bottom of the holon band at $\omega=0$.

The overall feature of $A^e({\bf k}, \omega)$ versus $\omega$ is shown in 
Fig. 5 at different ${\bf k}$'s along (0, 0) to ${\bf k}_0$. The whole 
energy range of the spectral function is about $16 J\approx 2.24 $$eV$
(for $J=0.14$$eV$). Here we have chosen $t_h=2J$ ($t=5J$) in (\ref{hh}) in obtaining
the holon spectral function $\rho_h$ which is calculated numerically by exactly
diagonalizing $H_h$ under random flux with $|\sum_{\Box}a_{ij}|\leq 0.86\pi$ in
the white noise limit. The sample size is $32\times 32$. 
\begin{figure}[t!]
\epsfxsize=8.0 cm
\centerline{\epsffile{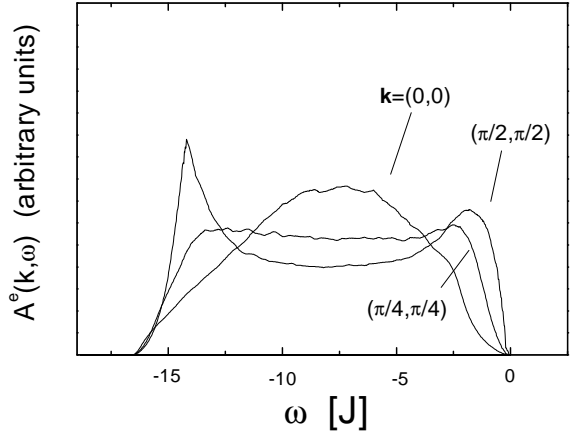}}
\vspace{2mm}
\caption{
The full shape of the spectral function calculated at ${\bf k}=(\pi/2,\pi/2)$,
$(\pi/4, \pi/4)$, and $(0,0)$. Note the whole energy range extends over
$16J\approx 2.24 $$eV$ (for $t=5J$, $J=0.14$$eV$).} 
\label{fig:5}
\end{figure}

Even though not delta-function-like, the spectral function do show peaks which
become sharper near both energy bottom and top as the momentum approaches ${\bf 
k}_0$. According to (\ref{hh}), the bottom and top of the holon band are near 
the momenta shifted away from (0,0) and ($\pi$,$\pi$), 
respectively, by ${\bf k}_0$, where the 
spectral function $\rho_h$ is sharpest in contrast to the broadest feature near 
the band center due to the random flux. On the other hand, the spinon spectrum 
$E^s_{\bf k}$ is maximum at ${\bf k}_0$ and
vanishes at momenta (0,0) and ($\pi$, $\pi$).
The convolution law of (\ref{spectral}) then combines
the contributions from the holon and spinon to result in Fig. 5. 
Note that the peaks
are asymmetric at the top and bottom of the band. The
effect of the spinon 
spectrum is most visible at lower binding energies, as will be further 
discussed below. 
A similar asymmetric structure in one-dimension has been 
found in Ref.\onlinecite{suzuura}.

Let us now focus on the structure of $A^e$ at low binding energies in comparison 
with ARPES measurements on Sr$_2$CuO$_2$Cl$_2$ \cite{wells,t'-t''} and 
Ca$_2$CuO$_2$Cl$_2$ \cite{ronning}.
In the left panel of Fig. 6, we plot $A^e$ within the
energy range of $8J\simeq 1.1$$eV$ at ${\bf k}$ positions along a line from 
(0,0) to ($\pi/2$, $\pi/2$) and then from ($\pi/2$, $\pi/2$) to ($\pi$, 0). The 
peak or edge position shows an isotropic dispersion along (0,0) to 
($\pi/2$, $\pi/2$) and from ($\pi/2$, $\pi/2$) to ($\pi$, 0), consistent 
with the experiments. Such a dispersion has a bandwidth about $2J$
and is clearly correlated with the spinon spectrum $E^s_{{\bf k}+{\bf k}_0}$.
The latter is marked by small bars at different ${\bf k}'s$ where
the position of the minimum $E^s_{{\bf k}+{\bf k}_0}$ at ${\bf k}_0$ is fixed
around $\omega=0.7J$ as a reference point. In the right panel of Fig. 6, a plot 
of $A^e$ along (0,0) and (0,$\pi$) is 
shown where the peak is replaced by an edge which looks dispersionless, 
also in good agreement with the experiment. Indeed, a very flat (only about 
$13\%$ change) $E^s_{{\bf k}+{\bf k}_0}$ along (0,0) to (0, $\pi$) is indicated 
in the figure by the small bars. In Fig. 7, the spinon 
spectrum $E^s_{{\bf k}+{\bf k}_0}$ fits the observed ARPES ``quasiparticle'' 
spectrum data reasonably well over the whole Brillouin zone along ${\bf k}_0$ 
($\Sigma$) to (0,0) ($\Gamma$) and (0,$\pi$) ($X$) (which are symmetric in our 
theory) with $J=0.14$$eV$.  Note that the low-energy scale is determined by
$J$ instead of $t$ and the features shown in Fig. 6 are not sensitive to
$t$. Thus, second - or further neighbor hopping terms are not
expected to play as important a role as they do in determining the
energy bottom, $\epsilon_{\bf k}$, shown in Fig. 1.
\begin{figure}[t!]
\epsfxsize=8.0 cm
\centerline{\epsffile{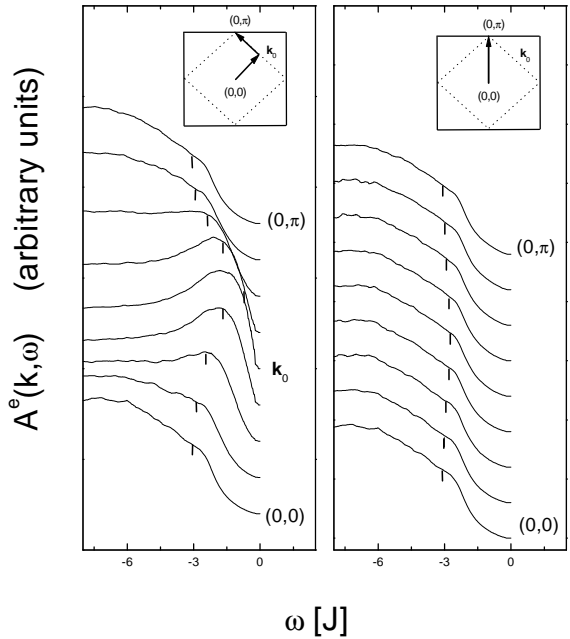}}
\vspace{2mm}
\caption{The spectral function at low binding energy along different momentum
scans. The small bars mark the spinon spectrum $E^s_{{\bf k}+{\bf k}_0}-\mu_0$
with $\mu_0=0.7J$.} 
\label{fig:6}
\end{figure}

The reason that the low energy peak or edge of the spectral function 
is correlated with the spinon
spectrum can be easily understood by noting the fact that $\rho_h({\bf 
k},\omega)$ becomes the sharpest near the bottom $\omega=0$ with ${\bf
k}\approx {\bf k}_0$ which contributes to $A^e$ in (\ref{ae}) at ${\bf
k}'\approx {\bf k}+ {\bf k}_0$ and $\omega\approx -E_{{\bf k}+{\bf k}_0}$.
In Fig. 8(a), the spectral function along (0,0) -- ${\bf k}_0$ is
shown when the maximal strength of $|\sum_{\Box}a_{ij}|$ is reduced from
$0.86\pi$ to $0.1\pi$. Here the correlation between the low-energy peak or 
edge positions and the spinon spectrum is even more apparent as the
holon spectral function $\rho_h$ becomes sharper near 
$\omega=0$ and ${\bf k}_0$. The evolution from a peak to an edge as 
${\bf k}$ moves away from ${\bf k}_0$ is because the spectral function of
the holon is quickly broadened away from the band bottom in the presence of
the random flux. Note that compared to the experiment the peak structure in
Fig. 8(a) is a bit too sharp which implies that the actual strength of
$|\sum_{\Box}a_{ij}|$ should be closer to our estimation used in Fig. 6.    
\begin{figure}[b!]
\epsfxsize=7.0 cm
\centerline{\epsffile{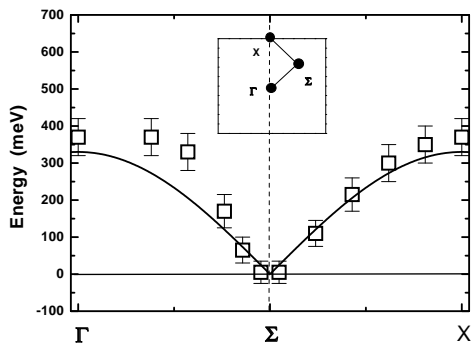}}
\vspace{2mm}
\caption{
The ``quasiparticle'' spectrum determined by the
ARPES \protect \cite{ronning} (open square) and by the spinon 
spectrum $E^s_{{\bf k}+{\bf k}_0}$ (solid curve).
} 
\label{fig:7}
\end{figure}

The line shape of the spectral function shown in Fig. 6 and Fig. 8(a) looks
strikingly similar to the ARPES measurements in Sr$_2$CuO$_2$Cl$_2$ 
and Ca$_2$CuO$_2$Cl$_2$ \cite{wells,t'-t'',ronning,add1,add2}. It reflects
essentially the {\it convolution} law of spinon and holon, i.e., the 
spin-charge separation. But we would like to point out that the simple 
convolution law is not enough. Here the {\it coherence} of the spinon in 
contrast to the {\em incoherent} holon is key to the characteristic features of the
line shape and the ``dispersion'' of the peak or edge structure. To illustrate
this point, we replace the coherent factor $v_{\bf k}^2$ in (\ref{ae}) by
a constant (i.e., $v_{\bf k}^2=1$), and re-calculate the spectral function
which is plotted in Fig. 8(b) along (0,0) -- ${\bf k}_0$. One sees that
the whole low-energy peak-edge feature in the left panel of Fig. 6 and 
Fig. 8(a) totally disappears in
Fig. 8(b). In this case, the summation in (\ref{ae}) smears out the peak
structure and $A^e$ is more or less like the density of states for the holon.
So the coherent factor $v^2$ in the spinon propagator is crucial for the
``quasiparticle'' peak to emerge in the spin-charge separation formulation
of the spectral function, which is the unique result of the {\it bosonic}
RVB description of spins. This explains why the correct line shape has not
been directly obtained in the slave-boson formalism, even though the 
prediction that the ``quasiparticle'' spectrum obtained in ARPES actually
corresponds to the spinon spectrum was first made\cite{laughlin} there.       
\begin{figure}[t!]
\epsfxsize=8.0 cm
\centerline{\epsffile{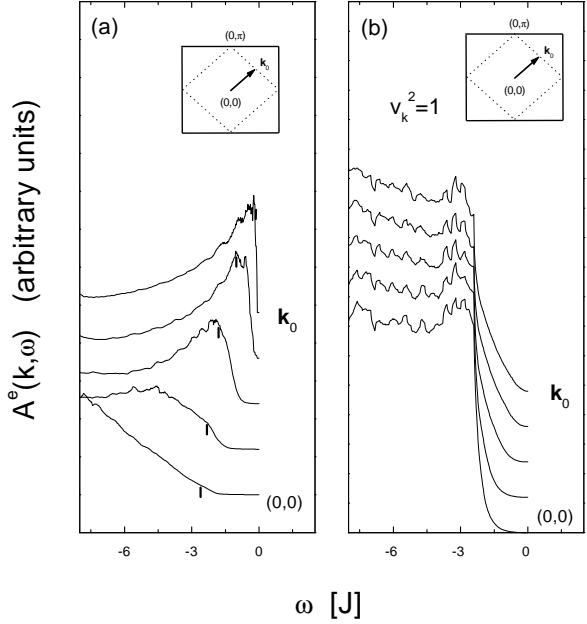}}
\vspace{2mm}
\caption{(a) The spectral function similar to the left panel of Fig. 6, but
with the random flux strength per plaquette reduced to $0.1\pi$ (see text). 
Small bars marked the same spinon spectrum as in Fig. 6 (but with $\mu_0=0.2J$);
(b) The line shape of the spectral function in the left panel of Fig. 6
is greatly changed if the spinon coherent factor $v_{\bf k}^2$ is replaced
by a constant $1$.} 
\label{fig:8}
\end{figure}

The incoherence of the holon is also an important factor as discussed above.
It is well-known that a particle must be generally localized, at least in the 
low-energy regime, in the presence of the random flux governed by $H_h$ in 
(\ref{hh}). We have checked the localization lengths corresponding to the
case of Fig. 6 using standard methods\cite{sheng}. 
The localization length quickly jumps 
from $4$-$15$ (in lattice units) to $36-50$ near the holon band edge which
covers an energy range related to the low-energy peaks of the spectral 
function in Fig. 6. As noted before, the localization length scale of the
holon is generally much larger than the maximum sample sizes ($<6\times 6$)
in numerical exact-diagonalization calculations\cite{leung}. 
Thus, numerical
calculations providing a ``metallic'' quasiparticle picture should be
irrelevant to the true picture at large length scales. 

Finally, to conclude our discussion of ARPES in the Mott insulator, we
turn to the experimental results of Ronning {\em at al.} \cite{ronning}. 
These authors identify a ``remnant Fermi surface'' in the insulator 
from ARPES measurements: they define the so-called relative momentum 
distribution
\begin{equation} 
n^r_k(\omega_0)=\int^{0}_{\omega_0}\frac{d\omega}{\pi}
A^e({\bf k},\omega)~~,
\end{equation}
with the cutoff energy $\omega_0=-0.5$$eV$, then identify ``$k_F$'' by
locating the position where $n^r_k$ drops suddenly. Then a ``remnant 
Fermi surface'' is found as the contour of steepest descent of 
$n^r_k$. Such a ``remnant Fermi surface'' roughly coincides with the
large Fermi surface expected for a free electron gas at half-filling 
concentration. It should be emphasized that this contour is {\em not} an 
equal energy contour. In fact, the peak (or edge) of the spectral function
disperses at momenta along the ``remnant Fermi surface'' as much as that 
shown in Fig. 7. 

So the ``remnant Fermi surface'' mainly indicates a strong momentum dependence 
of the energy-integrated spectral function, $n^r_k$. We believe this is 
due to the effect of the mobile hole on the spin background (at very short time scales). 
So far we have not considered the influence of the phase string effect
on the spin degrees of freedom. This is based on the fact that a single hole
does not affect the spin background at the thermodynamic scale. However,
the AF background {\em surrounding} the doped hole should be still strongly 
distorted by the hopping at $t\gg J$, which in turn will feed back on the hole 
itself through the phase string effect. Obviously, this happens at short length
and time-scales. Consequently, one has to go beyond the effective 
Hamiltonian (\ref{hh}) as well as the random flux treatment of the phase string 
effect. Not being within the main purview of this paper, we relegate the 
discussion of such effects to the Appendix.

\section{Conclusions and discussions}

Although antiferromagnetism in the Mott insulator has been well 
understood based on the bosonic RVB description, the problem of a single
hole doped into such a system is found
to be nontrivial. Unlike the spin-polaron picture in which the hole is 
predicted to behave like a Landau-Fermi quasiparticle carrying a finite-size
spin distortion around it, the real picture for the motion of the doped hole is 
that it always picks up a string of signs from the spin background during its
hopping, known as the phase string, which leads to the destruction of 
the ``coherence'' of the doped hole as a Landau-Fermi quasiparticle.

The phase string characterizes the effect of spin backflow during the
motion of the hole. The irreparable nature of phase string is intimately 
related to spin-charge separation: the hole cannot carry a precise spin-1/2 
quantum number with it. In contrast, if the holon and spinon were tightly 
confined together, the phase string effect would become trivial (reparable), as
is the case when one naively uses the $c$-operator to replace the spinon-holon 
decomposition in (\ref{ht}). Hence, the demonstration that the phase string
effect cannot be ``healed'' through spin superexchange interaction at low 
energies, presented in Sec. IIB and Refs.\onlinecite{string,string1}, 
points directly to spin-charge separation. It is also found that the spin
backflow, or equivalently, the phase string leads to mutual statistics
between the hole and spin degrees of freedom. That the backflow of spin
current could lead to mutual statistics and spin-charge separation was
also anticipated by Baskaran \cite{baskaran_92}.

In the one-hole case, the phase string effect mainly acts on the holon part,
causing its localization. But the spinon remains coherent in 
the bosonic RVB background. Consequently, during its propagation, 
a bare hole releases the coherent spinon constituent whose 
isotropic dispersion essentially 
determines the position of the low-energy peak or edge of the spectral 
function. On the other hand, spin-charge separation characterized by
the convolution law of the holon and spinon propagators is responsible for
the intrinsic broad feature of the spectral function at higher energy.
In the phase string formalism, the singular part of the phase string 
effect also decides the Fermi points ${\bf k}_0$ in the low-energy, long-time
limit as well as the large ``remnant Fermi surface'' structure in the 
high-energy, equal-time limit.
We thus obtain a systematic and natural explanation for the ARPES 
measurements within the framework of the $t-J$ model. Conversely, one may
say that the ARPES experiments have clearly presented evidence for 
spin-charge separation, as emphasized by Anderson.          

We find that the ``coherence''
of the spinon, reflected by the factor $v_{\bf k}^2$ in SBMFT, plays
an important role in determining the ARPES line shapes. A naive
convolution without this factor is not enough to explain the line shape
observed by ARPES. We also note that the strong 
temperature dependence of the ARPES line shapes
over a wide energy range \cite{kim1}, 
may be easily understood based on the temperature dependence of 
$v_{\bf k}^2$. Details will be presented elsewhere. Finally, we point out
that the temperature dependence of $v_{\bf k}^2$ is also crucial to get the
correct behavior of the relaxation rate ($1/T_1$) in nuclear magnetic resonance 
measurements, spin-spin correlation length, and other magnetic properties at
half-filling within the SBMFT.

The prediction that the holon as the charge carrier is localized by
the phase string effect is also consistent with the cuprate superconductors.
It is known that the phase with AFLRO (weakly doped regime)
is always an insulator. 
According to our theory, the localization is due to the intrinsic nature of
the doped Mott insulator instead of external reasons like the Anderson 
localization in the presence of impurities. 

Note that the localization of the hole is because the 
phase string is solely picked up by the hole, and the antiferromagnetic 
background remains the same as at half-filling, thermodynamically. At finite 
density of holes, both spin and charge degrees of freedom will 
be affected by the phase string, which can lead to a metallic (superconducting) 
phase\cite{string3} without AFLRO. We emphasize that
the irreparable phase string effect always exists, 
even in the metallic phase. This is because the phase string reflects
the competition between the hopping and 
superexchange interactions in the $t-J$ model, which has nothing to do with the 
existence of AFLRO. Thus, the phase string effect will persist at any
finite doping so long as short range antiferromagnetic correlations persist\cite{string1,string3}.


\acknowledgments We acknowledge useful 
discussions with A.~Ferraz, T.
~Tohyama, T.~K.~Lee, and especially C.~Kim and F.~Ronning who kindly provided
their experimental data. We also thank Barry Wells for discussions on
ARPES and useful correspondence. Work at Houston was
supported in part by Texas ARP No.3652707, the Robert A. Welch 
foundation, and TCSUH. V.N.M. is supported by NSF Grant DMR-9104873.

\appendix{\section{Short time scale effects and the ``Remnant Fermi surface''}}

The holon is localized by phase string in space. But at a short scale within
the localization length, the holon may behave like in the metallic phase at
finite doping where holons are mobile. It has been found\cite{string1,string3} 
that in the metallic phase the phase string effect will influence both the spin 
and charge degrees of freedom in
such a way that the correct decomposition form for the electron becomes
\begin{equation}\label{emutual}
c_{i\sigma}=h_i^{\dagger}\bar{b}_{i\sigma}e^{i\hat{\Theta}^{string}_{i\sigma}},
\end{equation}
where $h_i^{\dagger}$ and $\bar{b}_{i\sigma}$ are the ``true'' holon and spinon
operators and the phase shift field $e^{i\hat{\Theta}^{string}_{i\sigma}}$
precisely keeps track of the non-repairable phase string effect. In the 
following we will use this formalism to describe the local, high-energy 
regime in the one-hole case and show a nontrivial consequence of the phase
string effect.

In the one-hole case, the phase shift field can be written as 
$e^{i\hat{\Theta}^{string}_{i\sigma}}=(-\sigma)^i e^{\frac{i}{2}\Phi_{i}^{b}}$
in which 
\begin{equation}\label{phib}
\Phi^{b}_i= \sum_{l\neq i} 
\mbox{Im ln $(z_i-z_l)$}\left(\sum_{\alpha}\alpha n_{l\alpha}^b -1\right). 
\end{equation} 
with $z_i=x_i+iy_i$ representing the complex coordinate of a lattice site $i$
and $n^b_{l\alpha}$ denoting the spinon number at site $l$.

In order to understand $n^r_{k}$, let us first take $\omega_0$ to
$-\infty$. In this high-energy or equal-time limit, $n^r_k$ reduces to the 
momentum distribution $n_k$ which is given by
\begin{eqnarray}\label{nk}
n_{k}&=&\frac 1 N \sum_{ij}e^{i{\bf k}\cdot({\bf r}_j-{\bf r}_i)}
\langle c^{\dagger}_{j\sigma}c_{i\sigma}\rangle\nonumber\\
&=&\frac 1 N \sum_{ij} e^{i{\bf k}\cdot({\bf r}_j-{\bf r}_i)} 
\left\langle \bar{b}^{\dagger}_{j\sigma} h_j\left(e^{i \int_{\Gamma}d{\bf r}\cdot \hat{\bf A}^f({\bf r})}\right)h^{\dagger}_i(0)\bar{b}_{i\sigma}\right\rangle,
\end{eqnarray}
in which the following expression for the equal-time phase-string factor is
used: 
\begin{equation}\label{equalt1}
{\left.e^{i\frac 1 2 {\Phi}^b_{i}(t)}e^{-i\frac 1 2 {\Phi}^b_{j}(0)}
\right|}_{t\rightarrow 0^-}= e^{i \int_{\Gamma}d{\bf r}\cdot \hat{\bf A}^f({\bf r})}
\end{equation}
with
\begin{equation}\label{equalt2}
\hat{\bf A}^{f}({\bf r})=\frac 1 2 \sum_{l}\left[\sum_{\sigma} \sigma n^b_{l
\sigma}-1\right] \frac{{\bf \hat{z}}\times ({\bf r}-{\bf r}_l)}{|{\bf r}-
{\bf r}_l|^2}
\end{equation}
where the path $\Gamma$ connects two lattice sites $i$ and $j$ without
crossing other lattice sites. Clearly, in the equal-time limit, the momentum
structure will be mainly decided by the oscillation of the phase string factor
given in (\ref{equalt1}). 

For the one-hole case, $n_k$ in (\ref{nk}) is actually trivial and
featureless ($=1/2$). This is because $\langle  c^{\dagger}_{i\sigma} 
c_{j\sigma}\rangle= \delta_{ij}\langle \bar{b}_{i\sigma}^{\dagger}\bar{b}_{i\sigma}\rangle$ using the
non-double-occupancy constraint in which no propagation of the holon at
a finite distance can occur at strictly equal time on the half-filling
ground state. On the other hand, 
$n^r_k$ for a finite cutoff $\omega_0$ will involve a finite-time 
holon propagation over some finite distance. In this case, the phase
string factor in (\ref{equalt1}) will show up to determine the basic
feature while the rest of the propagator involving holon and spinon will
mainly give rise to a broadening in the momentum space depending on how far the
holon and spinon constituents can travel under the cutoff $\omega_0$.
At $\omega_0=-\infty$, the momentum broadening would reach infinity
(as only $i=j$ contributes) such that any momentum structure arising from
$e^{i \int_{\Gamma}d{\bf r}\cdot \hat{\bf A}^f({\bf r})}$ gets smeared out.
But at a large but finite $|\omega_0|$ the momentum structure due
to the phase string factor should show up. Here as an approximation the
equal-time phase string factor is still used so long as $|\omega_0|$ is 
sufficiently large.

In the spin channel, a line-integral expression similar to (\ref{equalt1}) also 
appears\cite{string4} in the spin-spin correlation function, leading to the \
so-called incommensurate antiferromagnetic peaks. Here we can borrow the 
similar method used in Ref.\onlinecite{string4} to manipulate the contribution
of the phase string factor (\ref{equalt1}) in $n^r_k$, which gives rise to
four ``incommensurate peaks'' at ($\pm \pi\kappa$, 0) and (0, $\pm \pi\kappa$) 
in momentum space with $\kappa \sim 1$\cite{remark}. If this oscillating factor 
solely decides the momentum structure of $n^r_k$ 
with the rest term in the propagator mainly contributing a broadening as
discussed above, then one gets a contour plot of $n^r_k$ in Fig. 9 in terms of
the superposition of aforementioned 
four peaks (with $\kappa=0.75$) at an arbitrary broadening (which is presumably 
controlled by the energy cutoff $\omega_0$: when $\omega_0\rightarrow -\infty$, 
the broadening in the momentum space should go to infinity such that 
$n_{\bf k}=1/2$). Note that the values of $n^r_{\bf k}$ here are only meant for
{\it relative} comparison. 

\begin{figure}[ht!]
\epsfxsize=7.0 cm
\centerline{\epsffile{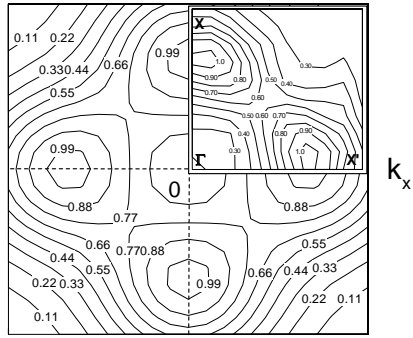}}
\vspace{2mm}
\caption{The contour plot of the electron momentum distribution 
exhibiting a ``remnant Fermi surface'' structure. The experimental 
data \protect \cite{ronning} are shown in the insert at the 
upper right corner.} 
\label{fig:9}
\end{figure}

We see that the overall feature is in qualitative
agreement with the experimental data\cite{ronning} whose one quarter portion is 
also shown in the inset of Fig. 9 for comparison. Here we remark that 
experimental data\cite{haffner,ronning1} of $n^r_k$ seem to be photon-energy
dependent, indicating the effect of the electron-photon matrix element,
and there also exits controversies\cite{haffner,ronning1} 
on whether the ``remnant 
Fermi surface'' defined near the sharpest drop of $n_k^r$ in ${\bf k}$-space
is meaningful or not. But in the present approach the momentum dependence
of $n^r_k$ does not represent any real Fermi surface structure of the Mott
antiferromagnet at half-filling. It only reflects the superposition of four
peaks near ($\pm \pi$, 0) and (0, $\pm\pi$), an enhancement entirely coming 
from the dynamic effect of the doped hole as the result of a 
careful handling of 
the phase string effect in the local, high-energy regime. As a prediction, if
experimentally $|\omega_0|$
is taken to sufficiently higher energy, the momentum dependence of $n^r_k$
should get weaker and weaker and eventually $n^r_k\rightarrow n_k=1/2$ at
$\omega_0\rightarrow -\infty$.


\begin{references}

\bibitem{wells} B.O. Wells, Z.-X. Shen, A. Matsuura, D.M. King, M.A. Kastner, M.Greven, and R.J. Birgeneau, Phys. Rev. Lett. {\bf 74}, 964 (1995).

\bibitem{t'-t''} C. Kim, P.J. White, Z.-X. Shen, T. Tohyama, Y. Shibata,
S. Maekawa, B.O. Wells, Y.J. Kim, R.J. Birgeneau, and M.A. Kastner, Phys. Rev. Lett. {\bf 80}, 4245 (1998).

\bibitem{ronning} F. Ronning, C. Kim, D.L. Feng, D.S. Marshall, A.G. Loeser, L.L. Miller, J.N. Eckstein, I. Bozovic, Z.-X. Shen, Science,{\bf 282}, 2067 (1998).

\bibitem{add1} S. LaRosa, I Vobornik, F. Zwick, H. Berger, M. Grioni, G. 
Margaritondo, R. J. Kelley, M. Onellion, and A. Chubukov, Phys. Rev. B{\bf 56}, 
R525 (1997).

\bibitem{add2} C. Durr, S. Legner, R. Hayn, S.V. Borisenko, Z. Hu, A. Theresiak,
M. Knupfer, M. S. Golden, J. Fink, F. Ronning, Z.-X. Shen, H. Eisaki, S. Uchida,
C. Janowitz, R. Muller, R.L. Johnson, K. Rossnagel, L. Kipp, and 
G. Reichardt, cond-mat/0007283.

\bibitem{laughlin} R.B. Laughlin, Phys. Rev. Lett. {\bf 79}, 1726 (1997);
J. Phys. Chem. Solids. {\bf 56}, 1627 (1995). 

\bibitem{scba} S. Schmitt-Rink {\it et al.} Phys. Rev. Lett. {\bf 60}, 2793 
(1988).

\bibitem{kane} C. L. Kane, P.A. Lee, and N. Read, Phys. Rev. B {\bf 39}, 6880 (1989).

\bibitem{scba2} G. Martinez and P. Horsch, {\it ibid}, B {\bf 44}, 317 (1991).

\bibitem{scba3} Z. Lui and E. Manousakis, {\it ibid}, B {\bf 44}, 2414 (1991).

\bibitem{diagonalization} see, E. Dagotto, Rev. Mod. Phys. {\bf 66}, 763 (1994)
and references therein.

\bibitem{leung} P.W. Leung and R.J. Gooding, Phys. Rev. B{\bf 52}, R15711 
(1995).

\bibitem{lee} T.K. Lee and C.T. Shih, Phys. Rev. B{\bf 55}, 5983 (1997).

\bibitem{t'} A. Nazarenko {\it et al.}, Phys. Rev. B {\bf 51}, 8676
(1995); R. Eder {\it et al.}, Phys. Rev. B{\bf 55}, R3414 (1997).

\bibitem{add} G. B. Martins, R. Eder, and E. Dagotto, Phys. Rev. B{\bf 60}, 
R3716 (1999). 

\bibitem{tohyama} T. Tohyama, Y. Shibata, S. Maekawa, Z.-X. Shen, and N. 
Nagaosa, J. Phys. Soc. Jpn, {\bf 69}, 9 (2000) and references therein. 

\bibitem{book} P. W. Anderson, {\em The theory of superconductivity in
the high-$T_c$ cuprates}, Princeton University Press, 1997. 

\bibitem{string} D. N. Sheng, Y. C. Chen, and Z. Y. Weng
Phys. Rev. Lett. {\bf 77}, 5102 (1996).

\bibitem{string1} Z. Y. Weng, D. N. Sheng, Y. C. Chen, and C. S. Ting, Phys. 
Rev. B{\bf 55}, 3894 (1997).

\bibitem{remark1} Here $(-\sigma)^i$ is introduced for convenience, which 
explicitly keeps track of the Marshall sign\cite{marshall}.

\bibitem{marshall} W. Marshall, Proc. Roy. Soc. (London) A{\bf 232}, 48
(1955).

\bibitem{lda_88} S. Liang, B. Doucot and P. W. Anderson, Phys. Rev.
Lett. {\bf 61}, 365 (1988).

\bibitem{chen}  Y. -C. Chen and K. Xiu, Phys. Lett. A{\bf 181}, 373 (1993).

\bibitem{aa} D.P. Arovas and A. Auerbach, Phys. Rev. B{\bf 38}, 316 (1988);
A. Auerbach, {\em Interacting Electrons and Quantum Magnetism} (Springer-Verlag,
New York, 1994).

\bibitem{remark2} For convenience, the magnetization here is to lie in the 
$x-y$ plane instead of along the $z$-axis as in Ref.\onlinecite{kane}.

\bibitem{remark3} Note that such a hole-position-dependent 
choice of the phase $\eta_{{\bf k}\sigma}$ is perfectly legitimate 
as in the sub-Hilbert-space in which the hole position is given, the 
commutation relations of ${b}$, ${b}^{\dagger}$ remain unchanged. On the other
hand, the spin-hole states are simply orthogonal if the hole is at different 
sites.

\bibitem{string3} Z. Y. Weng, D. N. Sheng, and C. S. Ting, Phys. Rev. Lett. {\bf 
80}, 5401 (1998); Phys. Rev. B {\bf 59}, 8943(1999).

\bibitem{suzuura} H. Suzuura and N. Nagaosa, Phys. Rev. B{\bf 56}, 3548 (1997).


\bibitem{sheng} see, D. N. Sheng and Z. Y. Weng, Europhys. Lett. {\bf 50}, 776 
(2000).

\bibitem{baskaran_92} G. Baskaran, Prog. Theor. Phys. Suppl. {\bf
107}, 49 (1992).

\bibitem{kim1} C. Kim, F. Ronning, A. Damascelli, D.L. Feng, Z.-X. Shen, B.O.
Wells, Y.J. Kim, R.J. Birgeneau, M.A. Kastner, L.L. Miller, H. Eisaki, and S. Uchida, preprint.

\bibitem{string4} Z. Y. Weng, D. N. Sheng, and C. S. Ting, Phys. Rev. B 
{\bf 59}, 11367(1999).

\bibitem{remark} Readers are referred to Ref.\onlinecite{string4} for the 
detail, where a similar phase-string factor appearing in the spin-spin 
correlation function is discussed which gives rise to incommensurate shifts by 
($\pm  2\pi\delta \bar{\kappa}$, 0), etc., from ($\pi$,$\pi$) with $\delta$ 
denoting the doping concentration and $\bar{\kappa}\sim 1$.


\bibitem{haffner} S. Haffner, D. M. Brammerier, C. G. Olson, L. L. Miller, and D. W. Lynch, cond-mat/0006366.

\bibitem{ronning1} F. Ronning, C. Kim, K. M. Shen, N. P. Armitage, A. Damascelli, D. H. Lu, Z.-X Shen, and L. L. Miller, cond-mat/0007252.

\end{references}
\end{document}